\documentclass{aa}
\usepackage{pifont}

\usepackage{graphicx}
\usepackage{txfonts}

\usepackage{tikz}
\usetikzlibrary{arrows,snakes,backgrounds}

\begin{document}

\title{Rayleigh scattering in the atmospheres of hot stars}
\author{J. Fi\v{s}\'{a}k
        \inst{1,2}
        \and
        J. Krti\v{c}ka\inst{1}
        \and
        D. Munzar
        \inst{1}
        \and
        J. Kub\'{a}t
        \inst{2}}
\institute{
Masarykova Univerzita,
P\v{r}{\'\i}rodov\v{e}decká fakulta,
Kotl\'{a}\v{r}sk\'{a} 2, Brno, Czech Republic
                \and
Astronomick\'y \'ustav, Akademie v\v{e}d \v{C}esk\'e republiky,
Fri\v{c}ova 298, 251 65 Ond\v{r}ejov, Czech Republic}

\abstract{
Rayleigh scattering is a result of an interaction of photons with bound electrons.
Rayleigh scattering is mostly neglected in calculations of hot star model atmospheres
because most of the hydrogen atoms are ionized and the heavier elements have a lower
abundance than hydrogen. In atmospheres of some chemically peculiar stars, helium overabundant
regions containing singly ionized helium are present and
Rayleigh scattering can be a significant opacity source.}
{We evaluate the contribution of
Rayleigh scattering by neutral hydrogen and singly ionized helium in the
atmospheres of hot stars with solar composition and in the atmospheres of
helium overabundant stars.}
{We computed several series of model atmospheres using the TLUSTY code
and emergent fluxes using the SYNSPEC code.
These models describe atmospheres of main sequence
B-type stars with different helium abundance. We used an existing grid of models 
for atmospheres with solar chemical composition and we calculated an additional grid
for helium-rich stars with $N$(He)/$N$(H)=10.}
{
Rayleigh scattering by neutral hydrogen can be neglected in atmospheres
of hot stars,
while
Rayleigh scattering by singly ionized helium can be a 
non-negligible
opacity source in some hot stars, especially in helium-rich stars.}
{}


\maketitle

\section{Introduction}

The interaction
of photons with bound electrons
is
very important
in stellar atmospheres.
Processes of excitation and ionization are very well
known and have been studied in detail.
However, photons may interact with bound electrons even  when
photon energy is not equal to the energy difference
between any bound levels. Rayleigh scattering is one of these processes.

Rayleigh scattering is a special kind of scattering process.
According to \cite{Loudon1983}, the scattering processes by bound electrons 
are divided into three cases:  fluorescence, Raman scattering,
and Rayleigh scattering. In the case of Rayleigh scattering an ion is excited by
a photon and transits to a virtual state. This state is very unstable and the
electron transits to the original state immediately.
If the initial state is different from the final state we call this process
Raman scattering.

Rayleigh scattering was studied  by 
John William Strutt \citep[Lord Rayleigh, see][]{rayleigh1870light}.
This effect causes the blue colour of the sky and the red colour of sunsets.
 Scattering centres are oxygen and nitrogen molecules because they 
have the largest abundance in the Earth's atmosphere. Rayleigh scattering
is also an important opacity source  in the atmospheres of other planets 
\citep{Buenzli2009}.

Rayleigh scattering by H$_2$ was detected in exoplanet atmosphere HD209458b
\citep{LecavelierdesEtangs2008} and it is suspected that  Rayleigh scattering is 
responsible for the  blue colour of exoplanet HD189733b 
\citep{LecavelierdesEtangs2008ii}. Rayleigh scattering is an
important opacity source in cool stars because of large neutral hydrogen
atom population \citep[see][]{Hubenyc2015}.

Scattering by bound electrons is often neglected in hot star atmospheres
because of the low abundance of Rayleigh scattering atoms.
First, both hydrogen and helium are ionized leaving only a small
fraction in the neutral state and second, abundances of the other
elements are low. In addition, the Rayleigh scattering cross section is
smaller for atoms with a higher atomic number.

However,
the population of singly ionized helium is much larger
than the population of neutral hydrogen in hot
main sequence stars with solar composition.
In addition, there are stars with helium overabundant
atmospheres, which means they have even larger
population of singly ionized helium.

The aim of this paper is to investigate the possible effect of Rayleigh
scattering by ionized helium on radiation emerging from model atmospheres of hot
stars.

\section{Rayleigh scattering}
\begin{figure}[t]
\centering
   \begin{tikzpicture}
        \draw[very thick] (0,0) -- (5,0) node [draw,fill=blue!20!,fill opacity=.5,text=black,near end,anchor=north,text ragged] {Ground state};
        \draw[thick] (1,1.3) -- (5,1.3) node [near end,anchor=north] {Excited state};
        \draw[thick] (1,2.5) -- (5,2.5) node [near end,anchor=north] {Excited state};
        \shade [ball color=green] (0.5,0) circle (3pt);
        \draw[->,purple,very thick] (0.5,0.1) .. controls (0.25,1) and (0.25,2) .. (0.9,2.9);
        \filldraw[green] (1,3) circle (3pt);
        \draw[->,red,very thick] (1.1,2.9) .. controls (2,2) and (2,1) .. (1.6,0.1);
        \shade[ball color=red] (1.5,0) circle (3pt);
\end{tikzpicture}
\caption{Scheme of the Rayleigh scattering process. 
         The bound electron is excited by an incoming photon to
         a virtual state which is unstable
         and the electron transits immediately to the original state.
        }
\label{pic:rozptylSchema}
\end{figure}
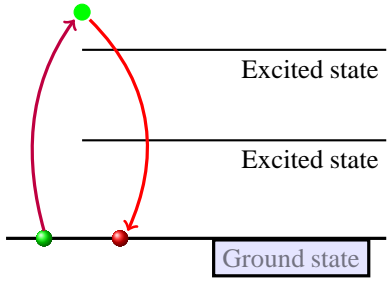

Rayleigh scattering is a
type of interaction of radiation 
with bound electrons. The energy of an interacting photon is not equal to
any energy difference between an excited state of the atom and the ground
state. An electron is excited to an unstable virtual state 
(in contrast to the line transitions), from which it transfers
immediately back to its original state.  This is shown in 
Fig.~\ref{pic:rozptylSchema}. 

The process of Rayleigh scattering is nicely described
by the time
dependent perturbation theory. Using an electric-dipole approximation and 
the \emph{Krammers-Heisenberg} formula for the differential
cross-section we obtain in SI units (see \citealt{Loudon1983})
\begin{multline}
\frac{\text{d}\sigma(\omega)}{\text{d}\Omega}=
 \frac{e^4\omega}{16c^4\pi^2\epsilon_0^2\hbar^2}\times\\
\Bigg|\sum_n\left(\frac{(\boldsymbol{\varepsilon}_{\boldsymbol{k}_\text{s}}\mathbf{D}_{1n})(\boldsymbol{\varepsilon}_{\boldsymbol{k}}\mathbf{D}_{n1})}
        {\omega_n-\omega}+\right.
 \left.
\frac{(\boldsymbol{\varepsilon}_{\boldsymbol{k}}\mathbf{D}_{1n})(\boldsymbol{\varepsilon}_{\boldsymbol{k}_\text{s}}\mathbf{D}_{n1})}
 {\omega_n+\omega}\right)\Bigg|^2,
 \label{eq:dcsRS}
\end{multline}
where $\mathbf{D}_{ji}=\big<j\big|r\big|i\big>$ is
the matrix element of the sum of the electron position vectors
(of the electron position vector for hydrogen-like atoms). Furthermore, 
$\boldsymbol{\varepsilon}_{\boldsymbol{k}}$ and
$\boldsymbol{\varepsilon}_{{\boldsymbol{k}}_\text{s}}$ are
the polarization vectors for the incoming photons with wave vector
$\boldsymbol{k}$ and for the scattered photons with wave vector $\boldsymbol{k}_\text{s}$,
respectively, and $\omega_n$ represents the angular frequency of the transition between the ground
state ($l$) and the $n$-th state
\begin{equation}
 \omega_n=\frac{E_n-E_l}{\hbar}.
\end{equation}
The sum runs over all intermediate states.
Here $e$ is the electron charge, $\varepsilon_0$
is the vacuum permittivity, $c$ the speed of light, and $m$ the electron mass.

After some calculations  described in \cite{Loudon1983} and \cite{Lee2004-01-21},
and after some mathematical arrangements 
(\citealt{Zettilic2009}, \citealt{Bethec2008}, and \citealt{Landau1982}) we
obtain the following equation for Rayleigh scattering cross section in the
case of one-electron atoms,
\begin{equation}
\sigma(\omega)=\frac{2e^4\omega^4}{3c^4\pi\epsilon_0^2\hbar^2}
\Bigg|\sum_n \frac{\big<n,1\big\|r\big\|1,0\big>^2}{\omega_n}
        \left(1+\frac{\omega^2}{\omega_n^2}+\frac{\omega^4}{\omega_n^4}+\dots\right)\Bigg|^2,
\label{eq:CSHeIIi}
\end{equation}
where $\big<n,1\big\|r\big\|1,0\big>$ is
the reduced matrix element, $\big\|1,0\big>$ represents the
1s state,  and $\big\|n,1\big>$ represents the np state, n
is the principal quantum number and p is 
the azimuthal quantum number
(both bound states and continuum states are 
included). The reduced matrix element is given by the Wigner-Eckart
theorem valid for any vector operator $\hat{A}_q$
(see Eq.~(7.326) in \citealt{Zettilic2009}):
\begin{equation}
        \label{eq:WEtheorem}
        \big<n',j',m'\big|\hat{A}_q\big|n,j,m\big> =
        \big<j,1;m,q\big|j',m'\big>\big<n',j'\big\|A\big\|n,j\big>.
\end{equation}
The first matrix element on the right side of
Eq.~\eqref{eq:WEtheorem} is the Clebsch-Gordan coefficient.
It is possible to  rewrite Eq.~(\ref{eq:CSHeIIi})
in terms of dimensionless angular
frequency $\tilde{\omega}_n$ and dimensionless position vector $\tilde{\vec{r}}$
(see \citealt{Lee2004-01-21}),
\begin{multline}
 \label{CSoRRwME}
\sigma(\omega)=\sigma_\text{e} \left(\frac{\omega}{\omega_{\mathcal{L}}}\right)^4
\Bigg|\sum_n \frac{\big<n,1\big\|\tilde{r}\big\|1,0\big>^2}{\tilde{\omega}_n}
\\\times
\left(1+\frac{1}{\tilde{\omega}^2_n}\frac{\omega^2}{\omega_{\mathcal{L}}^2}+
        \frac{1}{\tilde{\omega}^4_n}\frac{\omega^4}{\omega_{\mathcal{L}}^4}+\dots\right)\Bigg|^2,
\end{multline}
where
\begin{equation}
\tilde{\omega}_n=\frac{\omega_n}{\omega_\mathcal{L}},\,\tilde{r}=\frac{r}{a_0},
\end{equation}
the angular frequency of Lyman limit $\omega_{\mathcal{L}}$ and Bohr radius $a_0$ are
\begin{equation}
\omega_{\mathcal{L}} = \frac{e^4m}{32\hbar^3\epsilon_0^2\pi^2},\,
a_0 = \frac{4\pi \epsilon_0 \hbar^2}{me^2},
\end{equation}
and $\sigma_\text{e}$ is the Thomson scattering cross section
\begin{equation}
 \sigma_\text{e}=\frac{e^4}{6\varepsilon_0^2c^4m^2\pi}=0.665
\times
10^{-24}\,\text{cm}^2.
\end{equation}

\subsection{Cross section for Rayleigh scattering on hydrogen}
%

First, we consider only hydrogen 
atoms. In this case, the reduced matrix elements can be calculated
analytically. According to \cite{Lee2004-01-21},
the total cross section for Rayleigh scattering by neutral hydrogen is equal to
\begin{multline}
\frac{\sigma(f)}{1\,\text{cm}^2} \approx
             \frac{8.4147\times
             10^{-25}}{1\,\text{cm}^2}\,\left(\frac{\omega}{\omega_\mathcal{L}}\right)^4 +\\
             \frac{2.4855\times
             10^{-24}}{1\,\text{cm}^2}\,\left(\frac{\omega}{\omega_\mathcal{L}}\right)^6 +
             \frac{5.8604 \times
             10^{-24}} {1\,\text{cm}^2}\,\left(\frac{\omega}{\omega_\mathcal{L}}\right)^8+\\
             \frac{4.2736\times
           10^{-23}} {1\,\text{cm}^2}\,\left(\frac{\omega}{\omega_\mathcal{L}}\right)^{10}+
             \frac{2.6546\times
           10^{-23}} {1\,\text{cm}^2}\,\left(\frac{\omega}{\omega_\mathcal{L}}\right)^{12}+\\
             \frac{5.3933\times
           10^{-23}} {1\,\text{cm}^2}\,\left(\frac{\omega}{\omega_\mathcal{L}}\right)^{14}+
             \cdots,
 \label{eq:RSCSHI}
\end{multline}
where $f = \frac{\omega}{2\pi}$.

\subsection{Cross section for Rayleigh scattering on hydrogen-like atoms}
The radial wavefunction of hydrogen-like atoms is given by
\cite[see][]{Landauc1977}
\begin{multline}
R_{Z,nl}(r)=\frac{1}{(2l+1)!}\sqrt{\frac{(n+l)!}{2n(n-l-1)!}}\left(\frac{2Z}{na_0}\right)^{3/2}
\exp\left(-\frac{Zr}{na_0}\right)\left(\frac{2Zr}{na_0}\right)^l
\\\times
F\left(l+1-n,2l+2,\frac{2Zr}{na_0}\right),
 \label{eq:vlFceHlikeAZ}
\end{multline}
where $n$ is the principal quantum number, $l$ is the
orbital quantum number, $Z$ is the atomic number,
$a_0$ is
the Bohr radius, and $F\left(l+1-n,2l+2,\frac{2Zr}{na_0}\right)$
is the hypergeometric function. It can be seen that 
\begin{equation}
R_{Z,nl}(r)=R_{nl}(Zr)Z^{3/2},
\end{equation}
where $R_{nl}(r)$ are radial wavefunctions of hydrogen. For the matrix element we obtain
\begin{multline}
\big<\text{npm}\big|z\big|1\text{s}0\big>_Z = \int\limits_0^{2\pi}\text{d}\varphi\;
\int\limits_0^\pi\text{d}\theta\;\sin{\theta}\;\int\limits_0^\infty\text{d}r\;r^2
Z^{3/2}R_{10}(Zr)Y_{00}(\theta,\varphi)\\
\times
Z^{3/2} R_{n1}(Zr)Y_{1m}(\theta,\varphi) r \cos(\theta)=
\frac{1}{Z}\big<\text{npm}\big|z\big|1\text{s}0\big>_{Z=1}.
\end{multline}
Using this result, the Wigner-Eckart theorem, and the fact that the energies of the
excited states scale with the factor $Z^2$ 
we obtain
\begin{multline}
\sigma(\omega)=\sigma_\text{e} \left(\frac{\omega}{\omega_{\mathcal{L}}}\right)^4
\Bigg|\sum_n \frac{\big<\text{np}\big\|\overline{r}\big\|1\text{s}\big>^2}{Z^4\tilde{\omega}_n}
\\\times
\left(1+\frac{1}{Z^4\tilde{\omega}^2_n}\frac{\omega^2}{\omega_{\mathcal{L}}^2}+
        \frac{1}{Z^8\tilde{\omega}^4_n}\frac{\omega^4}{\omega_{\mathcal{L}}^4}+\dots\right)\Bigg|^2,
\end{multline}
where the meaning of the symbols is the same  as
in Eq.~(\ref{CSoRRwME}). Using the same approach as in \cite{Lee2004-01-21},
we get for the Rayleigh scattering cross section by hydrogen-like atoms
\begin{multline}
 \label{eq:scatHlikeZ}
\frac{\sigma(f)}{1\,\text{cm}^2} \approx
             \frac{1}{Z^8}\frac{5.5758\times
             10^{-25}}{1\,\text{cm}^2}\,\left(\frac{\omega}{\omega_\mathcal{L}}\right)^4 +\\
             \frac{1}{Z^{12}}\frac{1.8567\times
             10^{-24}}{1\,\text{cm}^2}\,\left(\frac{\omega}{\omega_\mathcal{L}}\right)^6 +\\
             \frac{1}{Z^{16}}\frac{4.9480\times
             10^{-24}} {1\,\text{cm}^2}\,\left(\frac{\omega}{\omega_\mathcal{L}}\right)^8+\cdots.
\end{multline}
Finally, for singly ionized helium ($Z=2$) we obtain
\begin{multline}
\frac{\sigma(f)}{1\,\text{cm}^2} \approx \frac{3.2870 \times
             10^{-27}}{1\,\text{cm}^2}\,\left(\frac{\omega}{\omega_\mathcal{L}}\right)^4 +\\
             \frac{6.0682\times
             10^{-28}}{1\,\text{cm}^2}\,\left(\frac{\omega}{\omega_\mathcal{L}}\right)^6 +
             \frac{8.9423\times
             10^{-29}} {1\,\text{cm}^2}\,\left(\frac{\omega}{\omega_\mathcal{L}}\right)^8+\cdots.
 \label{eq:CSHeII}
\end{multline}
It can be easily seen that the expression on the right-hand side of Eq. (\ref{eq:dcsRS})
adapted to the hydrogen atom and also the corresponding expression on the right-hand
side of Eq.~\eqref{eq:CSHeIIi} converges only for $\omega/\omega_\mathcal{L}\ll 1$.
We have checked that for $\lambda>1500\,\AA$ the cross
section for the hydrogen atom can be approximated within a precision of about 5\%
by the sum of the first six terms of
the series that are given explicitly in Eq.~\eqref{eq:RSCSHI}. The cross section
for the singly ionized helium can be similarly approximated for $\lambda>1000\,\AA$  by
the sum of the first three terms of the corresponding series that are given explicitly in
Eq.~\eqref{eq:CSHeII}. The results 
were obtained using the approximations mentioned above.

\subsection{Rayleigh scattering opacity}
%

We can estimate the opacity due to the Rayleigh scattering
using the formula
$$
 \chi \sim n_i \sigma_i,
$$
where $n_i$ is number density of particles $i$
(which stands for \ion{H}{i} or \ion{He}{ii}) and $\sigma_i$ is
the corresponding Rayleigh scattering cross section.
\begin{table}[t]
 \centering
 \caption{Opacity of
Rayleigh scattering at
$\tau_\text{Ross}\approx 1$
for a model atmosphere with $T_\text{eff}=23\,000$K
together with its comparison with the opacities of the Balmer
continuum and electron scattering. The cross section
is calculated for frequency $2.4 \times 10^{15}\,\text{s}^{-1}$ where
the flux is largest (according to Wien's displacement law).
In this table $n$ stands for particle number density, $\sigma$ for
the Rayleigh scattering cross section, and $\chi$ for the opacity.
Lower subscripts \ion{He}{II} and \ion{H}{I} of
cross section and opacity
stand for Rayleigh scattering
and $e$ for free electrons. $\chi_\ion{H}{i2}$ is the Balmer continuum opacity
and $n_\ion{H}{i2}$ number of hydrogen atoms with $n=2$.
}
 \begin{tabular}{c c}
 \hline
  solar chemical composition & helium-rich stars\\
 \hline
  $n_\ion{He}{ii} \approx 10^4 n_\ion{H}{i}$ & $n_\ion{He}{ii} \approx 10^6 n_\ion{H}{i}$ \\
 \multicolumn{2}{c}{${\sigma_\ion{He}{ii} \approx 10^{-3} \sigma_\ion{H}{i}}$} \\
  $\chi_\ion{He}{ii} \approx 10 \chi_\ion{H}{i}$ & $\chi_\ion{He}{ii} \approx 10^3 \chi_\ion{H}{i}$ \\
 \hline
 ${n_\ion{He}{ii} \approx 10^{-1}n_e}$ &
 ${n_\ion{He}{ii} \approx n_e}$ \\
 \multicolumn{2}{c}{${\sigma_\ion{He}{ii} \approx 10^{-3}\sigma_e}$}  \\
 ${\chi_\ion{He}{ii} \approx 10^{-4}\chi_e}$ &
 ${\chi_\ion{He}{ii} \approx 10^{-3}\chi_e}$ \\
 \hline
 ${n_\ion{He}{ii} \approx 10^{5}n_\ion{H}{i2}}$ &
 ${n_\ion{He}{ii} \approx 10^6 n_\ion{H}{i2}}$ \\
 \multicolumn{2}{c}{${\sigma_\ion{He}{ii} \approx 10^{-9}\sigma_\ion{H}{i2}}$}  \\
 ${\chi_\ion{He}{ii} \approx 10^{-4}\chi_\ion{H}{i2}}$ &
 ${\chi_\ion{He}{ii} \approx 10^{-2}\chi_\ion{H}{i2}}$ \\
 \hline
 ${\chi_\ion{He}{ii} \approx 10^{-4}\left(\chi_\ion{H}{i2}+\chi_e\right)}$ &
 ${\chi_\ion{He}{ii} \approx 10^{-3}\left(\chi_\ion{H}{i2}+\chi_e\right)}$ \\
 \hline
 \hline
 \end{tabular}
 \label{tab:opEst}
\end{table}
In  Table~\ref{tab:opEst} we list relative estimates of the opacity
caused by Rayleigh scattering for hydrogen and singly ionized helium.
The number densities of particles were taken from computed models by the
TLUSTY code and
cross sections from \citet{Hubenyc2015} and Eq.~\eqref{eq:CSHeII}.
From this table it follows that the opacity of Rayleigh scattering by singly
ionized helium is approximately ten times larger than the opacity of Rayleigh scattering by
neutral hydrogen in atmospheres with solar chemical composition. It is approximately
one thousand times larger in helium-rich atmospheres ($N(\text{He})=10\,N(\text{H})$).
Consequently, Rayleigh scattering by singly ionized helium can be a more significant
opacity source than Rayleigh scattering by neutral hydrogen.
For stars with enhanced helium abundance, like helium-rich chemically
peculiar stars, Rayleigh scattering by singly ionized helium may be an even
stronger opacity source.
However, the opacity due to Rayleigh scattering by singly ionized helium is
significantly smaller than the opacity due to hydrogen bound-free transitions
and free electrons, which dominate the continuum opacity in hot star
atmospheres.

\section{Computation of model atmospheres}

For our analysis we used several series of NLTE models for 
atmospheres of main sequence
B-type stars with solar chemical composition and
of helium-rich stars with enhanced helium abundance  $N(\text{He})/N(\text{H})=10$ 
($N$ denotes number density). We used model atmospheres for effective temperatures
between 15 and 30\,kK and surface gravity $\log g =4$ (see Table~\ref{tab:param}).
\begin{table}[t]
 \centering
 \caption{Solar chemical composition (relative number of particles, adopted from
        \citealt {Grevesse1998}) and chemical composition of helium-rich stars.}
 \begin{tabular}{p{3cm} p{1cm} p{1cm} p{1cm}}
     & H & He & other \\
     \hline
     solar composition & 91.2\,\% & 8.7\,\% & 0.1\,\% \\
He-rich
stars & 9.08\,\%& 90.82\,\%& 0.1\,\% 
 \end{tabular}
 \label{tab:chemCom}
\end{table}
\begin{table}[t]
 \centering
 \caption{Parameters used for computed models of atmospheres.}
 \begin{tabular}{c c c}
  \hline
  $\mathbf{T_\text{eff}}$ & \textbf{solar ab.} & $\mathbf{N(\text{He})/N(\text{H})=10}$ \\
  \hline
  15000 & \ding{52}& \ding{53} \\
  16000 & \ding{52}& \ding{53} \\
  17000 & \ding{52}& \ding{53} \\
  18000 & \ding{52}& \ding{53} \\
  19000 & \ding{52}& \ding{52} \\
  20000 & \ding{52}& \ding{52} \\
  21000 & \ding{52}& \ding{52} \\
  22000 & \ding{52}& \ding{52} \\
  23000 & \ding{52}& \ding{52} \\
  24000 & \ding{52}& \ding{52} \\
  25000 & \ding{52}& \ding{52} \\
  26000 & \ding{52}& \ding{52} \\
  27000 & \ding{52}& \ding{52} \\
  28000 & \ding{52}& \ding{52} \\
  29000 & \ding{52}& \ding{52} \\
  30000 & \ding{52}& \ding{52} \\
  \hline
 \end{tabular}
 \label{tab:param}
\end{table}

For atmospheres with solar chemical composition we used
models from the grid \citep{Lanz2003,Lanz2007}
calculated with the TLUSTY\footnote{\url{http://nova.astro.umd.edu/}}
code. Since there are no model grids available for helium overabundant 
stars, we calculated them ourselves using the TLUSTY 
code \citep{Hubeny1988,Hubeny1995}. To be consistent with the precalculated grid of
\cite{Lanz2003,Lanz2007}, for all other elements except hydrogen and
helium we used the solar composition adopted by \cite{Grevesse1998}.

In the following step we used the
SYNSPEC\footnote{\url{http://nova.astro.umd.edu/Synspec49/synspec.html}}
code to calculate the emergent flux. In this step
we added a new opacity source -- Rayleigh scattering by singly ionized
helium -- to the SYNSPEC source code.
This new part of the code is very similar to the part that computes
Rayleigh scattering by neutral hydrogen.
Using the code we computed synthetic spectra with and without Rayleigh scattering
included.

\begin{figure}[t]
 \includegraphics[scale=.75]{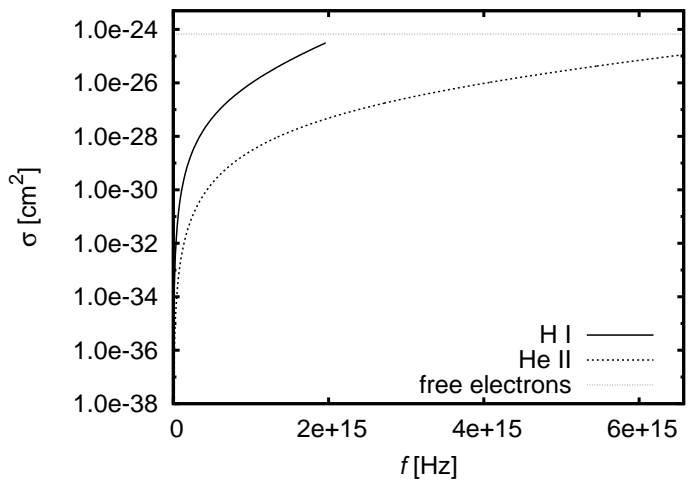}
 \caption{Adopted values of
	Rayleigh scattering cross section by neutral hydrogen
        (full line) and singly ionized helium (dotted line) in comparison with
        scattering cross section by free electrons (dashed line).}
 \label{fig:RSCS}
\end{figure}

For
easier
comparison of the computed synthetic spectra,
we 
convolved
the computed emergent fluxes with
the Gaussian function
$g(\lambda)$
$$
  (H*g)(\lambda)=\frac{1}{\sqrt{2\pi}\sigma}\int\text{d}\lambda'\;H(\lambda')
 \exp\left(-\frac{(\lambda-\lambda')^2}{2\sigma^2} \right),
$$
where $\sigma=10\,\AA$ is the width of the Gaussian function, $H(\lambda)$ is the emergent flux,
and $\lambda$ denotes the wavelength. Convolution makes the emergent flux smoother, and we are mainly
interested in continuum changes. We compared convolved emergent fluxes
between models with and without Rayleigh scattering. We plotted the difference between emergent spectra
in magnitudes given by
\begin{equation}
 \Delta m = -2.5\log\frac{H_+}{H_-},
\end{equation}
where $H_+$ is the emergent flux with Rayleigh scattering included and $H_-$ is the flux
without Rayleigh scattering included.

\section{Influence of Rayleigh scattering on emergent stellar radiation}

The computed difference in emergent fluxes depends on the atmospheric chemical
composition and the scattering cross section of the given ion. The difference also depends  on
the stellar effective temperature because  ionization and excitation
equilibria depend on temperature.

\subsection{Scattering by neutral hydrogen}
\begin{figure}
\includegraphics[scale=.65]{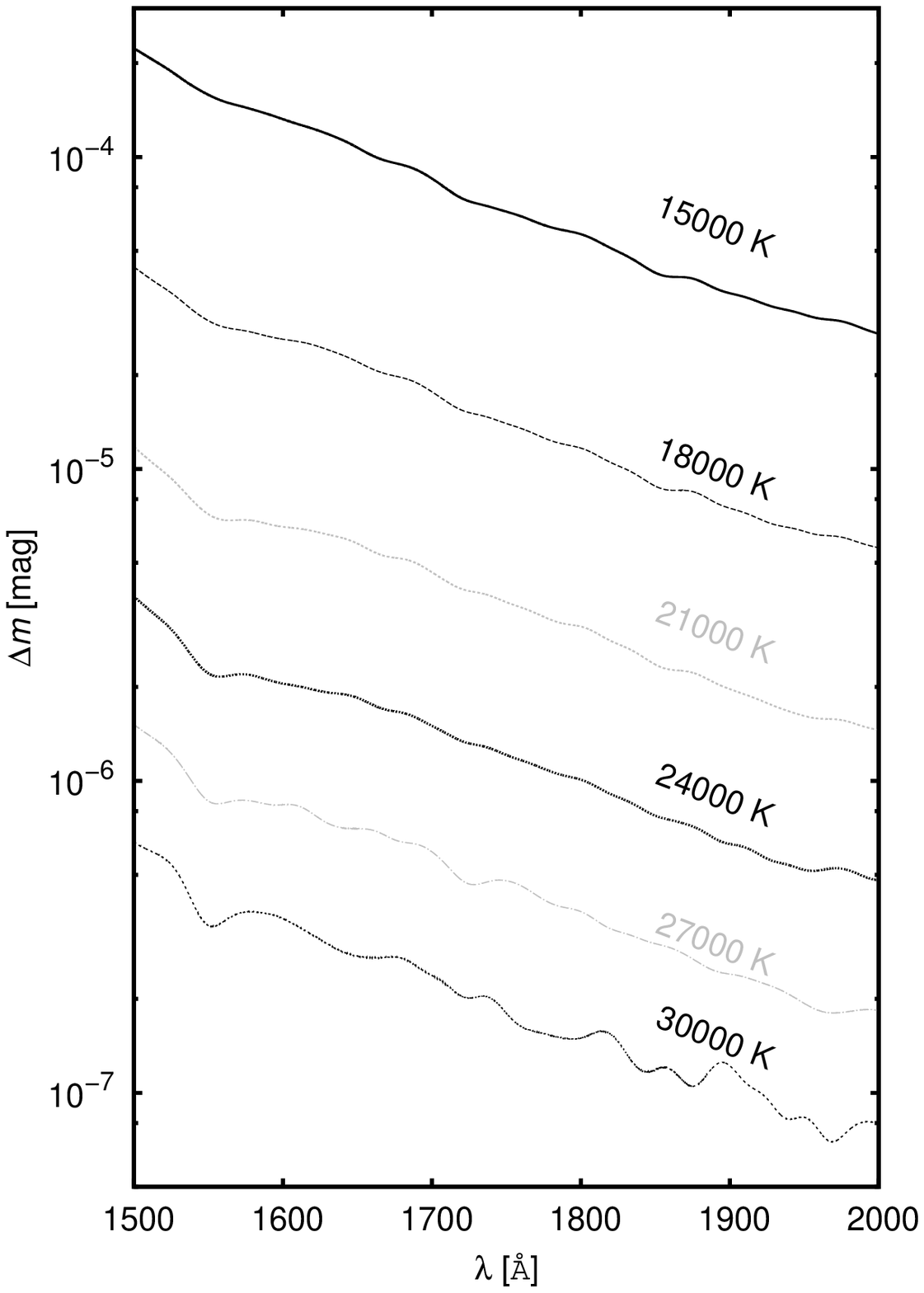}
                \caption{Dependence of difference of emergent fluxes with and without Rayleigh
                scattering by \ion{H}{I} included for given effective
                temperatures and solar chemical composition.}
\label{fig:diffNCrHI}
\end{figure}

We expected very small changes when adding  Rayleigh scattering by neutral hydrogen.
Rayleigh scattering by neutral hydrogen is included in the SYNSPEC code as an optional
opacity source. Our calculations confirmed its weaker influence on emergent fluxes
for temperatures larger than 21000\,K (in comparison to scattering by \ion{He}{II};
see Fig.~\ref{fig:diffNCrHI}).
The differences  decrease rapidly with the increasing effective
temperature of the star. 

\subsection{Scattering by singly ionized helium}

\begin{figure}
\includegraphics[scale=.65]{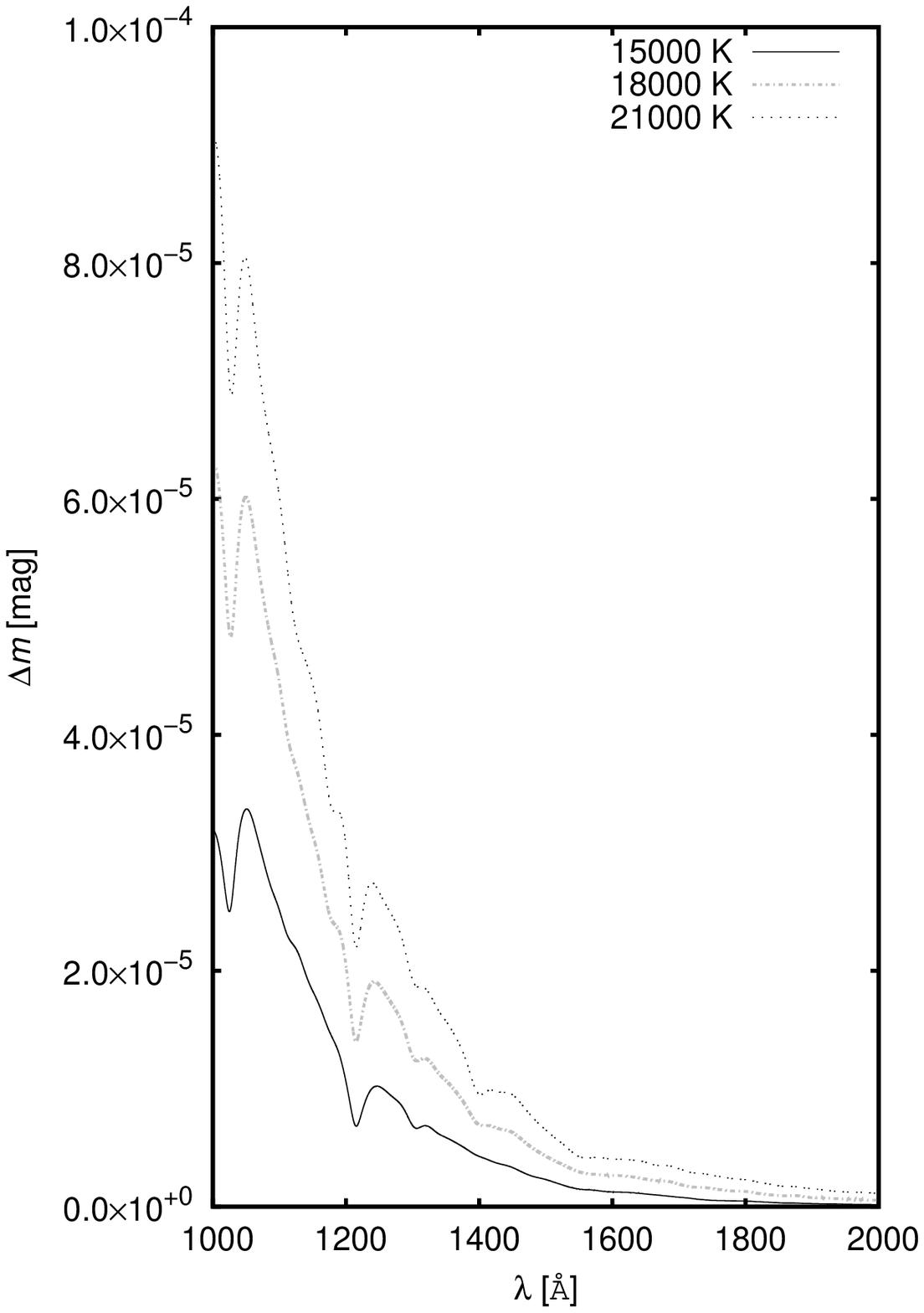}
\includegraphics[scale=.65]{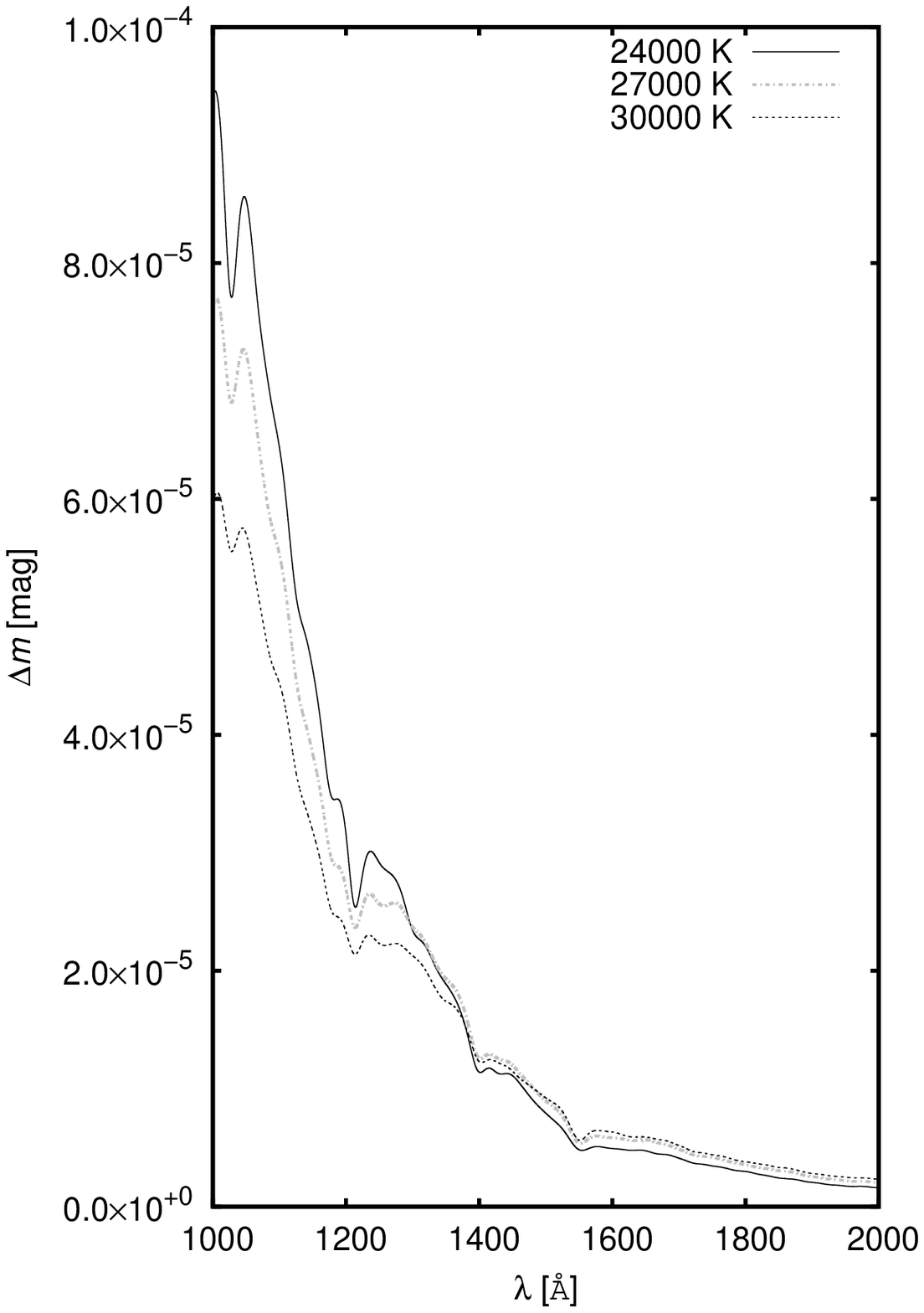}
\caption{Dependence of difference of emergent fluxes with and without Rayleigh
                scattering by \ion{He}{II} included for given effective
                temperatures and solar chemical composition.}
\label{fig:diffNCrHIHeII}
\end{figure}

In atmospheres with solar chemical composition 
the opacity due to the Rayleigh scattering by singly ionized helium is
larger than the opacity due to the Rayleigh scattering by hydrogen
for the temperatures in the interval (21, 30)\,kK.
This causes greater differences between emergent fluxes, which also depend
 on the stellar effective temperatures (see
Fig.~\ref{fig:diffNCrHIHeII}, where fluxes for selected models are plotted).
For lower effective temperatures the magnitude difference increases
with increasing effective temperature, but for effective temperatures
higher than about 25~kK the magnitude difference decreases. This behaviour
is caused by the population of singly ionized helium, which increases with
increasing effective temperature for cooler stars because of neutral helium
ionization; instead, in stars with effective temperatures higher than 25~kK,
the process of ionization of singly ionized helium becomes more
significant and lowers the \ion{He}{ii} abundance.

The Rayleigh scattering cross section is larger in the ultraviolet region
(see  Eq.~\ref{eq:CSHeII}). Consequently, Rayleigh scattering redistributes
the flux from the short-wavelength part of the spectrum to longer
wavelengths and the magnitude difference decreases with increasing wavelength.

\begin{figure}
\includegraphics[scale=.65]{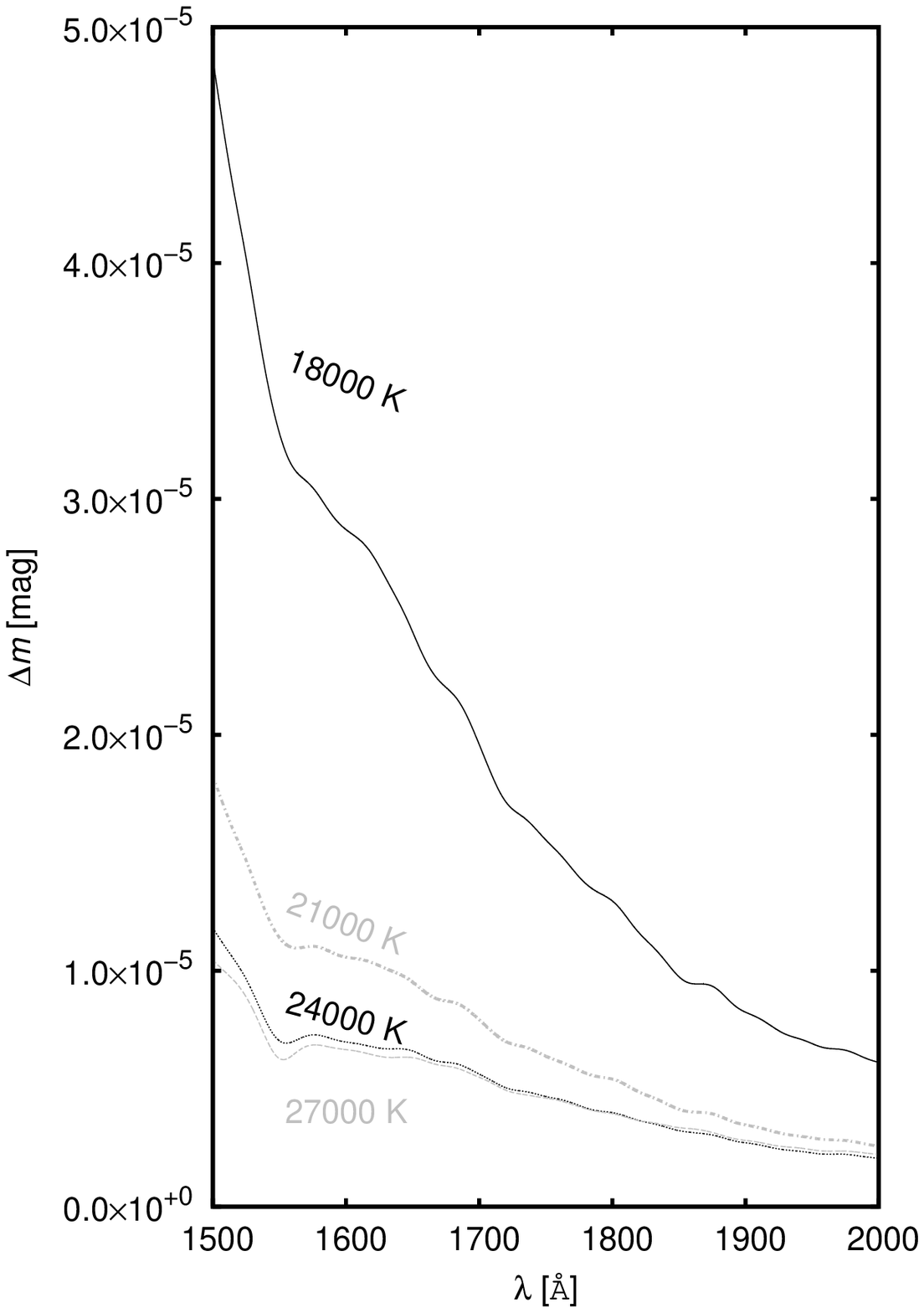}
\caption{Dependence of difference of emergent fluxes with and without Rayleigh
                scattering by \ion{H}{I} and \ion{He}{II} included for given effective
                temperatures and solar chemical composition.}
\label{fig:diffNCrHeII}
\end{figure}

We show the contribution of Rayleigh scattering by \ion{H}{I} and \ion{He}{II}
(see Fig~\ref{fig:diffNCrHeII}). We plotted differences only for temperatures
in the interval (18000, 24000)\,K. In this interval the contributions of both
processes are of the same order.

\begin{figure}[b]
\includegraphics[scale=.65]{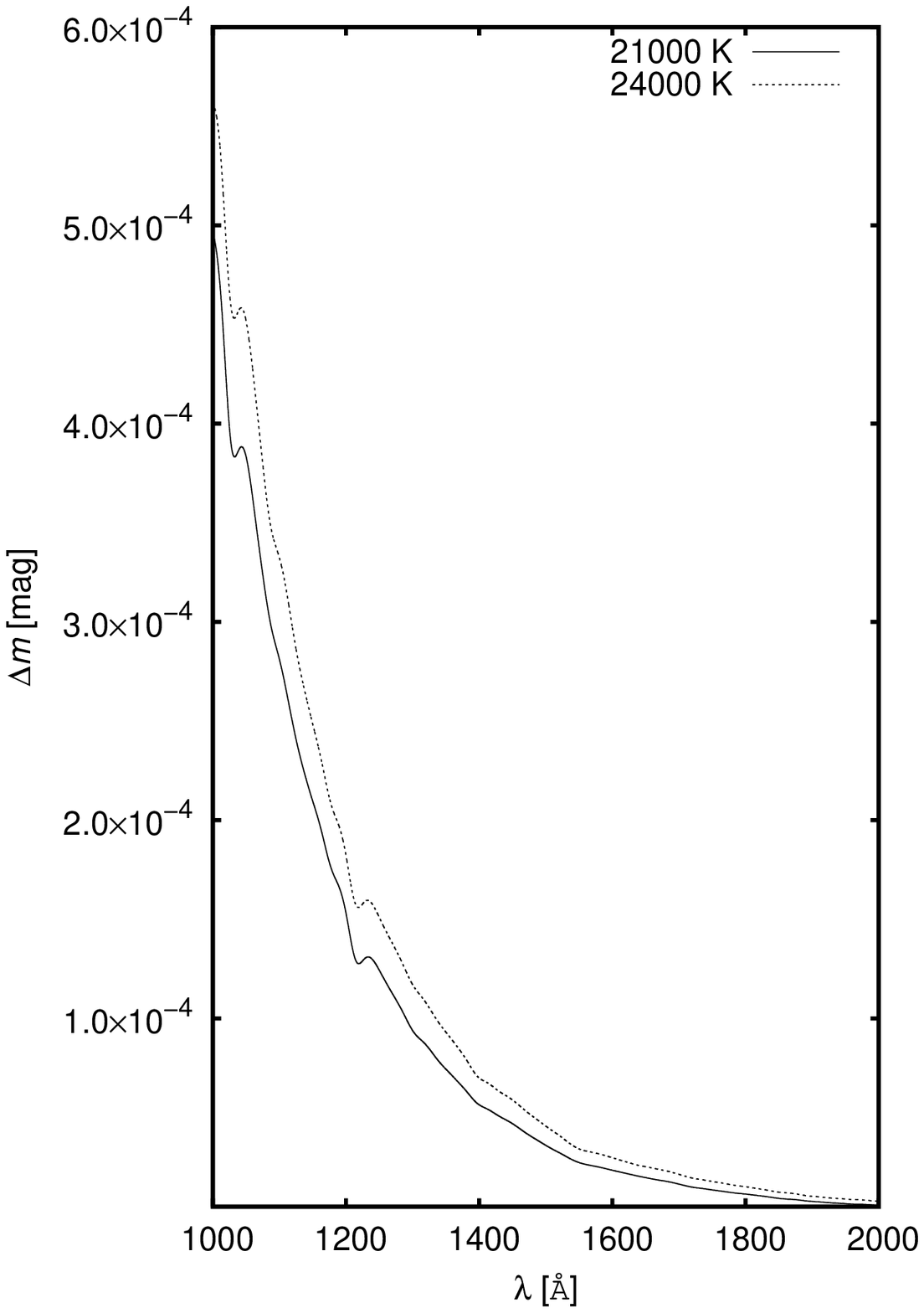}
\includegraphics[scale=.65]{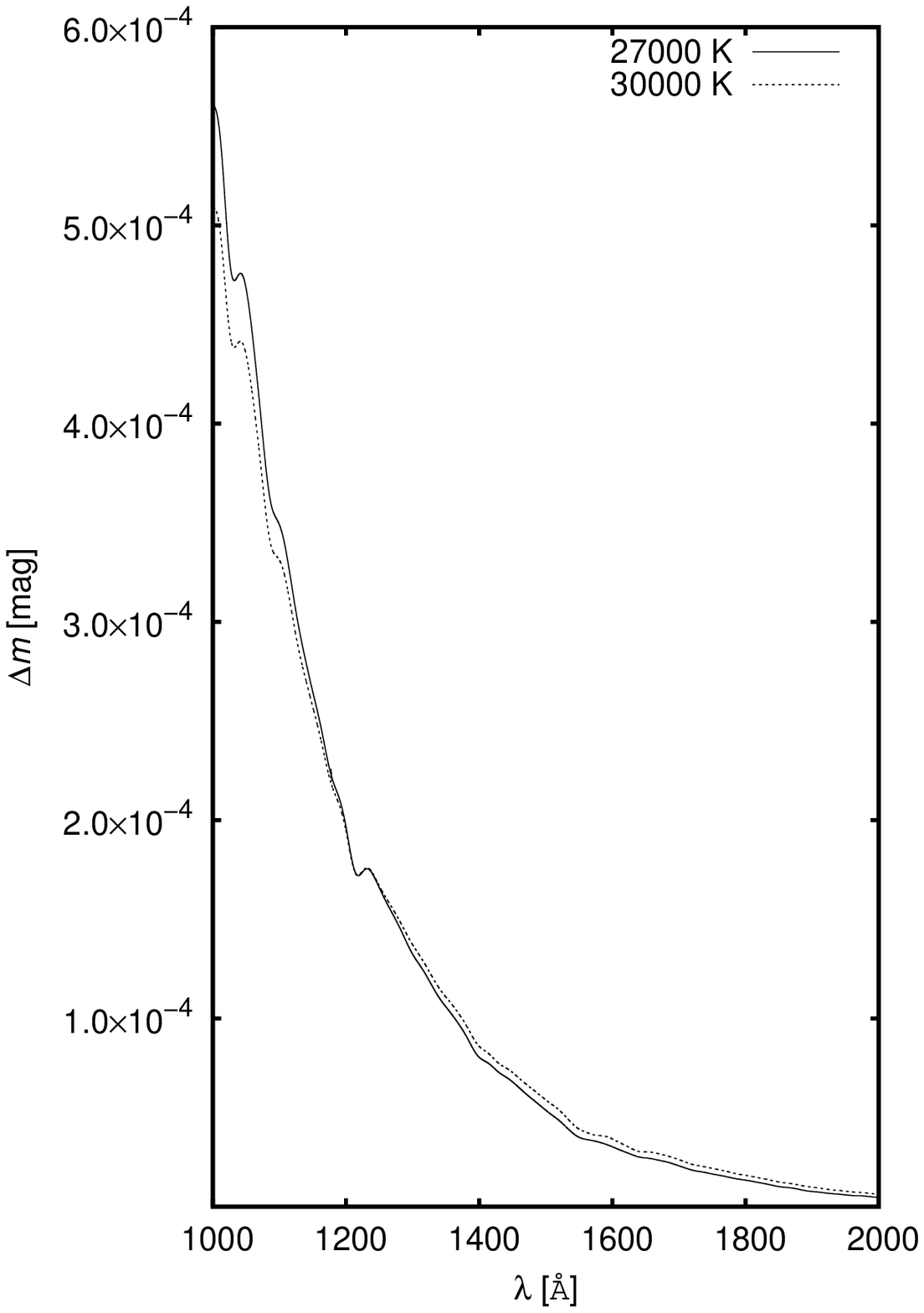}
                \caption{Dependence of difference of emergent fluxes with and without Rayleigh
                scattering by \ion{He}{II} included for given effective
                temperatures and helium overabundant chemical composition.}
\label{fig:CPHEflux}
\end{figure}

The magnitude difference is roughly  an order of magnitude
larger in the case of helium-rich stars (see Fig.~\ref{fig:CPHEflux}).
For the case of inhomogeneous surface distribution of helium (like in CP stars)
the synthetic spectra or lightcurves with these distributions must be computed.
For such stars the changes due to the Rayleigh scattering are larger than a
typical precision of space-based photometry.

\section{Conclusions}
In this paper we
study the effect of Rayleigh scattering on the emergent fluxes in hot stars.
We
derived the Rayleigh scattering cross section for \ion{He}{ii} (see Eq.~\ref{eq:CSHeII}).
We included this expression
into the SYNSPEC source code. Input models were computed using the TLUSTY code (for helium overabundant stars)
or we used a model grid 
of B stars (for stars with stellar chemical composition). We compared computed emergent fluxes with and without Rayleigh scattering.

The computed differences are relatively small. For stars with solar composition
the magnitude difference is of the order of $10^{-5}$, while for helium
overabundant stars the difference is  an order of magnitude higher.
The stellar winds are accelerated mostly in the UV region; we can thus
expect some influence on the line driving.
However, this analysis cannot be done with a static
model atmosphere code like TLUSTY (SYNSPEC). Rayleigh scattering is also a
potentially important source of opacity
 in other stars with a large abundance of helium, for example white
dwarfs, Wolf-Rayet stars, and subdwarfs.
\begin{acknowledgements}
This research was supported by  grant
GA\,\v{C}R 14-02385S. Access to computing and storage facilities owned
by parties and projects contributing to the National Grid Infrastructure
MetaCentrum, provided under the programme ``Projects of Large Infrastructure
for Research, Development, and Innovations'' (LM2010005), is greatly appreciated. 
\end{acknowledgements}
\bibliographystyle{aa}
\bibliography{./Literatura}
\end{document}